\documentclass{llncs}

\usepackage{xspace,amsmath,url}
\usepackage{color}
\usepackage{booktabs}
\usepackage[utf8]{inputenc}
\usepackage{comment}
\usepackage{cite}
\usepackage{url}
\usepackage{graphicx}

\DeclareMathSymbol{\mlq}{\mathord}{operators}{``}
\DeclareMathSymbol{\mrq}{\mathord}{operators}{`'}

\newcommand{\exclude}[1]{}

\newcommand{\argmax}[1]{\underset{#1}{\operatorname{arg}\,\operatorname{max}}\;}

\newcounter{lnoc}
\newenvironment{sdalgorithm}[1]{%
\hrule height 0.8pt \vspace{0.6ex} \small#1\vspace{0.6ex}\hrule height 0.5pt \vspace{-2.0ex}
\setcounter{lnoc}{0}
\small
\begin{tabbing}
000000\=XXI\=XXI\=XXI\=XXI\=XXI\=XXI\=\kill
}{%
\end{tabbing}
\vspace{-2.0ex}\hrule height 0.8pt\vspace{0ex}}
\newcommand{\lno}[1][0]{{\footnotesize\sffamily 
\ifnum#1=0
\stepcounter{lnoc} 
\ifnum\thelnoc<10
\phantom0%
\fi
\thelnoc
\else
\thelnoc.#1
\fi
}\>}

\newcommand{\pcfor}{{\bfseries for~}}

\newcommand{\pcdownto}{{\bfseries downto~}}
\newcommand{\pcdo}{{\bfseries do~}}
\newcommand{\pcif}{{\bfseries if~}}
\newcommand{\pcthen}{{\bfseries then~}}
\newcommand{\pcelse}{{\bfseries else~}}

\newcommand{\pcwhile}{{\bfseries while~}}

\newcommand{\pcand}{{\bfseries and~}}

\newcommand{\pcreturn}{{\bfseries return~}}

\usepackage[hidelinks]{hyperref}
\hypersetup
{
    pdfauthor={Szymon Grabowski and Aleksander Cis{\l}ak},
    pdftitle={A bloated {FM}-index reducing the number of cache misses during the search},
    pdfborder={0},
    urlcolor=black,
    colorlinks=false
}

\begin{document}

\title{A bloated {FM}-index reducing the number of cache misses during the search}

\author{Szymon Grabowski$^\dag$ and Aleksander Cis{\l}ak$^\ddag$}
\institute{
$^\dag$ Lodz University of Technology, Institute of Applied Computer Science,\\
  Al.\ Politechniki 11, 90--924 {\L}\'od\'z, Poland, 
  \email{sgrabow@kis.p.lodz.pl}\\
$^\ddag$ Warsaw University of Technology, Faculty of Mathematics and Information Science,\\
ul. Koszykowa 75, 00--662 Warsaw, Poland, \email{a.cislak@mini.pw.edu.pl}\\
}

\maketitle

\begin{abstract}
The FM-index is a well-known compressed full-text index, based on the Burrows--Wheeler transform (BWT).
During a pattern search, the BWT sequence is accessed at ``random'' locations, 
which is cache-unfriendly.
In this paper, we are interested in speeding up the FM-index by working on $q$-grams 
rather than individual characters, at the cost of using more space.
The first presented variant is related to an inverted index on $q$-grams, 
yet the occurrence lists in our solution are in the sorted suffix order rather than text order in a traditional inverted index.
This variant obtains $O(m/|CL| + \log n \log m)$ cache misses in the worst case, 
where $n$ and $m$ are the text and pattern lengths, respectively, and $|CL|$ 
is the CPU cache line size, in symbols (typically 64 in modern hardware).
This index is often several times faster than the fastest known FM-indexes 
(especially for long patterns), yet the space requirements are enormous, 
$O(n\log^2 n)$ bits in theory and about 80$n$--95$n$ bytes in practice.
For this reason, we dub our approach {\em FM-bloated}.
The second presented variant requires $O(n\log n)$ bits of space.
\end{abstract}

\section{Introduction}
The FM-index~\cite{FM00,FM05} invented by Ferragina and Manzini is a well-known 
compressed data structure based on the Burrows--Wheeler transform (BWT) which can answer full-text queries.
Given a range of text suffixes starting a sequence $S$, 
it allows to find quickly all text suffixes starting with $cS$, 
for any alphabet symbol $c$, 
using a mechanism called LF-mapping.
An inherent property of LF-mapping is however non-local access 
to the BWT sequence, which typically results in $\Omega(m)$ CPU cache misses 
during the search for a pattern of length $m$, even for a small alphabet.
(Handling a large alphabet of size $\sigma$ gives an extra time factor, 
e.g., $\log\sigma$ in popular wavelet tree based implementations; 
see~\cite{MN07,Nav13,BN14} for more details and known tradeoffs.)

The problem of cache misses during the FM-index backward search 
was identified as the main performance limiter 
by Chac{\'{o}}n et al.~\cite{CMEH13}, 
who proposed to perform the LF-mapping with several symbols at a time 
(in practice, at most 3, for the 4-symbol DNA alphabet for which 
the scheme was only described).
This solution allowed, for example, to improve the search speed by a factor 
of 1.5 for the price of doubling the index space.

In this paper we address the problem of cache misses during 
a pattern search ($count$ query) in a way related to 
the Chac{\'{o}}n et al.~solution. 
We also work on $q$-grams, yet the algorithmic details are different.
We propose two variants, one with $O(m/|CL| + \log n \log m)$ misses 
in the worst case, but also $O(n\log^2 n)$ bits of space, 
which makes it hardly practical, 
the other with $O(m/|CL| + (m/q') \log n)$ misses on average 
and $O(n\log n)$ bits of space, where $q' \geq 1$ is some parameter 
and $|CL|$ is the cache line size (nowadays, typically 64 bytes).
While in practice characters over an alphabet of size $\sigma$ are ``embedded'' 
in chunks of $b \geq \log\sigma$ bits (it often holds that $b = 8$, i.e.~the chunks 
are bytes), it is convenient to assume that $b = \Theta(\log\sigma)$, and we use this assumption henceforth.
Note that the well-known String B-tree data structure~\cite{FerraginaG99} 
achieves $O((m / B) \log_{B} n)$ I/Os 
(plus $(occ/B)$ I/Os for reporting the matches), which is optimal 
in the I/O comparison model.
The parameter $B$, the block size (expressed in the number of items stored in it), 
corresponds to our cache line size.

We use a standard notation throughout the paper, namely $S[i \ldots j]$ for a sequence $S$ and a pair of indexes $i$ and $j$ 
denotes a concatenation of symbols $S[i], S[i+1], \ldots, S[j]$.
In particular, $S[i \ldots j]$ is an empty string when $i > j$.
The index is built for the text $T[1 \ldots n]$, 
allowing to search for the pattern $P[1 \ldots m]$.
All logarithms are in base 2.

A preliminary version of the first presented variant was introduced in the Master thesis of the second author~\cite{cislak2015full}.

\section{An index with superlinear space}
\label{Sec:superlinear}
In accordance with the title we present here 
a truly bloated index (later referred to as FM-bloated), namely requiring $O(n\log^2 n)$ bits of space.
Like in any FM-index, we sort all suffixes of $T$, 
but instead of taking the single symbols standing before the suffixes 
in the sorted order (which form the BWT of $T$), we take all such $q$-grams 
for all relevant values $q$ being the power of 2.
Namely, for each suffix $T[i+1 \ldots n]$ we take the $q$-grams:
$T[i]$ (1-gram), $T[i-1 \ldots i]$ (2-gram), 
$T[i-3 \ldots i]$ (4-gram), ...,
$T[i-2^{\lfloor \log i \rfloor}+1 \ldots i]$.
Let $Q$ denote a collection of all such $q$-grams for all $i$.
For each distinct item $x$ from $Q$ we create a list $L_x$ of its occurrences 
in the sorted suffix order (later we use the term {\em suffix array order} 
or simply {\em SA order}).
This resembles an inverted index on $q$-grams, yet the main difference 
is that the elements of the lists are arranged in SA order 
rather than text order in a traditional inverted index.

For a given pattern $P[1 \ldots m]$, we start the LF-mapping with its 
longest suffix of length being a power of 2, namely, 
$P[m-2^{\lfloor \log m \rfloor}+1 \ldots m]$.
The following backward steps deal with the remaining prefix of $P$ 
in a similar way.
Note that the number of LF-mapping steps is the number of 1s in the 
binary representation of $m$, which is $O(\log m)$.
Figure~\ref{alg:FMbloated1} illustrates this approach.

In our representation, each LF-mapping step translates to performing 
two predecessor queries on a list $L_x$.
A na{\"i}ve solution is a binary search with $O(\log n)$ worst-case time (or even a linear search, which may turn out to be faster for very short lists), 
yet the predecessor query can essentially be handled in $O(\log\log n)$ 
time~\cite{Patrascu08a} using an y-fast trie~\cite{Willard83}.
In terms of cache misses we can however resort to the the optimal search 
in the I/O {\em comparison} model~\cite{vitter2008algorithms}, 
obtaining $O(\log_{|CL|\log\sigma/\log n}(n))$ cache misses.

We also test a variant in which the values of $q$ are Fibonacci numbers 
instead of powers of 2, with the starting values of 1 and 2.
For example, if $m = 20$, the pattern is decomposed to substrings of length 
13, 5, and 2.

\begin{figure}[h!]
\begin{sdalgorithm}{Bloated1-Count-Occs($n$, $C$, $L$, $P$, $m$)}
\lno	$ i \gets m$\\
\lno	$ q' \gets 2^{\lfloor \log i \rfloor}$\\
\lno	$\mathit{sp} \gets 1$; $\mathit{ep} \gets n$ \\
\lno	\pcwhile $((\mathit{sp} \leq \mathit{ep})$~\pcand~$(i\geq 1)$ \pcdo \\
\lno	\> $qGram \gets P[i-q'+1 \ldots i]$ \\
\lno	\> $sp \gets C[qGram]+\mathit{Occ}(L_{qGram}, \mathit{sp}-1)+1$ \\
\lno	\> $ep \gets C[qGram]+\mathit{Occ}(L_{qGram}, \mathit{ep})$ \\
\lno	\> $i \gets i-q'$ \\
\lno	\> $ q' \gets 2^{\lfloor \log i \rfloor}$\\
\lno	\pcif $(\mathit{ep} < \mathit{sp})$ \pcthen \pcreturn ``not found'' \pcelse \pcreturn ``found ($\mathit{ep}-\mathit{sp}+1$) occs''
\end{sdalgorithm}
\caption{Counting the number of occurrences of pattern $P$ in $T$ with 
the superlinear-space FM-bloated variant.
The function $Occ(\cdot)$ is defined as
$Occ(L_x, pos) = \argmax{i\in\{1,\ldots,|L_x|\}}(L_x[i] \leq pos)$.
To simplify the notation, we assume that $\log 0 = 0$ (line 9).}
\label{alg:FMbloated1}
\end{figure}

\section{An index with linear space}

\begin{figure}[h!]
\begin{sdalgorithm}{Bloated2-Count-Occs($n$, $C$, $L$, $P$, $m$, $q$, $p$)}
\lno	find $1 \leq i_1 < \ldots < i_{m'} \leq m$ such that \\
\> \> \> each $P[i_j]$ is a start position of a $(q, p)$-minimizer in $P$\\
\lno	$\mathit{sp} \gets 1$; $\mathit{ep} \gets n$ \\
\lno	\pcfor $j \gets m$ \pcdownto $i_{m'}$ \pcdo \ \ \ /* pattern suffix */ \\
\lno	\> $sp \gets C[j]+\mathit{Occ}(L_{P[j]}, \mathit{sp}-1)+1$ \\
\lno	\> $ep \gets C[j]+\mathit{Occ}(L_{P[j]}, \mathit{ep})$ \\
\lno	\pcfor $j \gets m'-1$ \pcdownto $1$ \pcdo \\
\lno	\> $qGram \gets P[i_j \ldots i_{j+1}-1]$ \\
\lno	\> $sp \gets C[qGram]+\mathit{Occ}(L_{qGram}, \mathit{sp}-1)+1$ \\
\lno	\> $ep \gets C[qGram]+\mathit{Occ}(L_{qGram}, \mathit{ep})$ \\
\lno	\pcfor $j \gets i_1-1$ \pcdownto $1$ \pcdo \ \ \ /* pattern prefix */ \\
\lno	\> $sp \gets C[j]+\mathit{Occ}(L_{P[j]}, \mathit{sp}-1)+1$ \\
\lno	\> $ep \gets C[j]+\mathit{Occ}(L_{P[j]}, \mathit{ep})$ \\
\lno	\pcif $(\mathit{ep} < \mathit{sp})$ \pcthen \pcreturn ``not found'' \pcelse \pcreturn ``found ($\mathit{ep}-\mathit{sp}+1$) occs''
\end{sdalgorithm}
\caption{Counting the number of occurrences of pattern $P$ in $T$ with 
the linear-space FM-bloated variant.
The function $Occ(\cdot)$ is defined as
$Occ(L_x, pos) = \argmax{i\in\{1,\ldots,|L_x|\}}(L_x[i] \leq pos)$.}
\label{alg:FMbloated2}
\end{figure}

In this variant we use at most two $q$-grams preceding 
any $T[i+1 \ldots n]$ suffix, hence reducing the overall space 
to $O(n\log n)$ bits.

The solution makes use of so-called minimizers, 
proposed in 2004 by Roberts et al.~\cite{RHHMY04} 
and seemingly first overlooked in the bioinformatics (or string matching) 
community, only to be revived in the last years 
(e.g.,~\cite{LKHYYS13,WS14,ChikhiLJSM15,DeorowiczKGD15,GrabowskiDR15}). 
A minimizer for a sequence $S$ of length $r$ is the lexicographically 
smallest of its all $(r-p+1)$ $p$-grams (or $p$-mers, in the term commonly 
used in bioinformatics); usually it is assumed that $p \ll r$.
For a simple example, note that two DNA sequencing reads 
with a large overlap are likely to share the same minimizer, 
so they can be clustered together.
That is, the smallest $p$-mer may be the identifier of the bucket into 
which the read is then dispatched.

In our variant, 
we pass a sliding window of length $q$ over $T$ and 
calculate the lexicographically smallest 
substring of length $p$ in each window (i.e., its minimizer).
Ties are resolved in favor of the leftmost of the smallest substrings.
The positions of minimizers are start positions of the $q$-grams into 
which the text is partitioned 
(note it resembles the recently proposed SamSAMi index, a sampled suffix 
array on minimizers~\cite{GR15}).
The values of $q$ and $p$, $p \leq q$, are construction-time parameters.

Let us assume that $1 \leq i_1 < i_2 < \ldots < i_{n'} \leq n-p+1$, $n' \leq n$, 
are the start positions of the obtained $q$-grams in $T$.
We add to $Q$ all $n$ unigrams $T[i]$, $1 \leq i \leq n$, 
and for each position $i_j$, $2 \leq j < n'$, 
also the $q$-gram $T[i_{j-1} \ldots i_j - 1]$ assuming its length is greater 
than 1 (as all unigrams from $T$ are already added).

For a given pattern $P[1 \ldots m]$ we pass a sliding window over it 
finding the start positions of all its minimizers, which partition 
the pattern into several (variable-length) $q$-grams.
If $P$ occurs in $T$, all those $q$-grams must belong to $Q$.
The following LF-mapping, together with the discussion on representation 
of the $L_x$ lists, is identical as in the previous variant.
The LF-mapping on the pattern boundaries works on (zero or more) unigrams.
Figure~\ref{alg:FMbloated2} illustrates this approach.

Assume that the average distance between the start positions of the found 
minimizers in $T$ (or in $P$) is about $(q-p+1)/2$; let us denote it with $q'$.
This gives us $O(m/q')$ LF-mappings on average, if $q' \leq m$.
As we don't know $m$ at the index construction time, we cannot choose 
a proper $q'$ (or rather $q$ and $p$) to guarantee $o(m)$ LF-mappings on average.
This can be helped with increasing the index space by 
a factor of $\Theta(\log\log n)$.
To this end, we look for minimizers over $T$ (and $P$, in the query time) 
several times; 
first for some constant $q_1$ (and naturally a constant $p_1$, $p_1 < q_1$), 
then for $q_2 = q_1^2$, then for $q_3 = q_2^2$, and so on, as long as the current 
window size is not greater than $n$.
Assuming that the average distances $q'_i$ between the start positions of 
the minimizers also grow approximately quadratically, 
there exists $q'_j$ such that 
$q'_j \geq \sqrt{m} / 2$ and $q'_j \leq m/4$.
This gives 
$O(m^{1-1/2}) = O(m^{1/2})$ LF-mappings per pattern on average.

\section{Implementation and experimental results}

\begin{figure}[h!]
    \centering
    \includegraphics[scale=0.30]{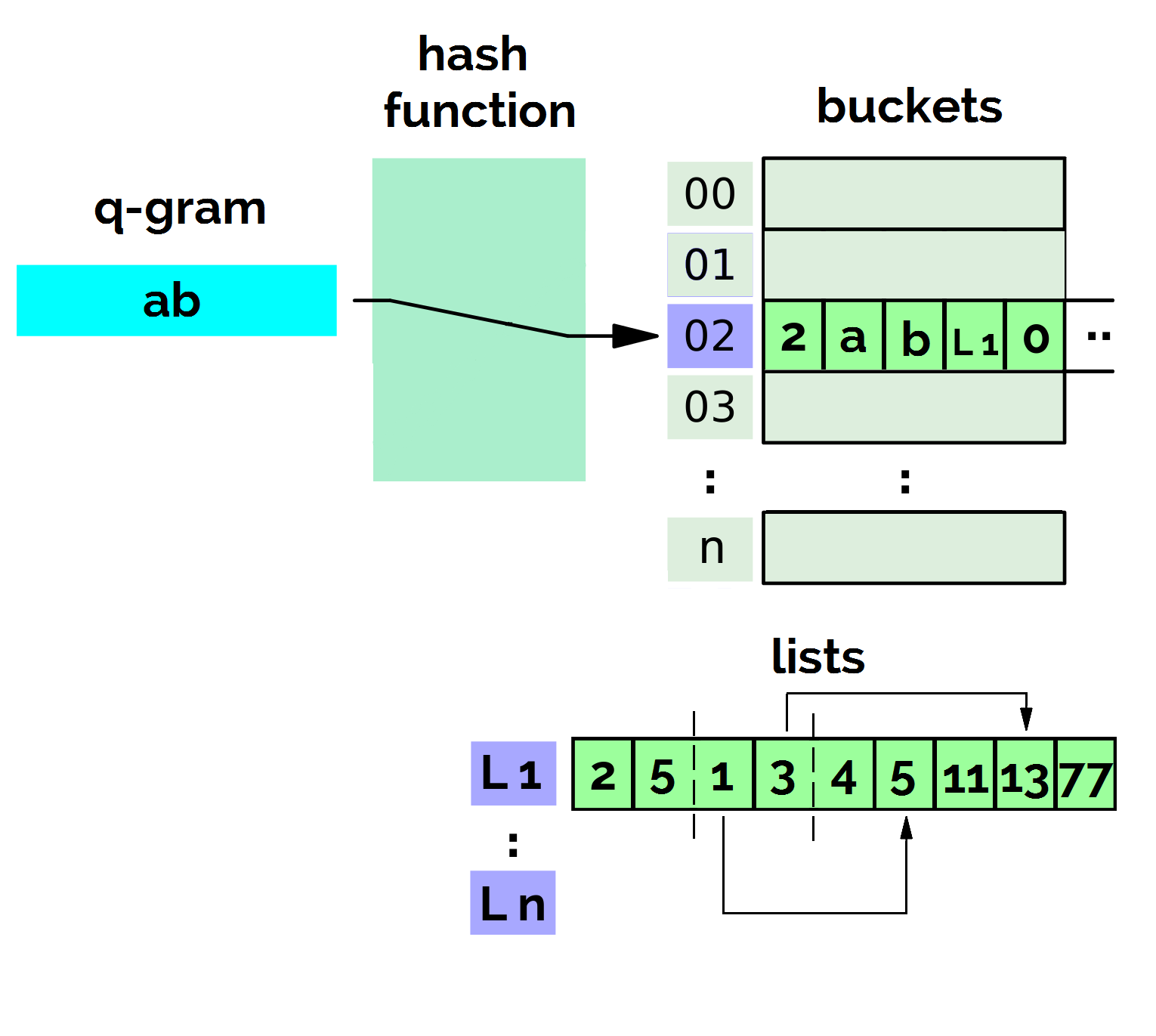}

    \caption{Layout of the hash table used by the FM-bloated index for full-text indexing. We can see the insertion of a $q$-gram ``ab''. The words are shown for illustrative purposes, and in the actual implementation only pointers to the original text are stored. The buckets contain $q$-gram sizes (stored as 8-bit characters) as well as pointers to their respective lists. In this case, the value of the count table $C[\mlq\mlq ab \mrq\mrq]$ is equal to 2 (first value on the list), and there are 5 positions (the number of positions is stored as the second value on the list) which will be used for the calculation of $Occ(\cdot)$, namely 4, 5, 11, 13, and 77 (all arbitrary values). Moreover, we can see the quick access (QA) list of length 2 whose beginning and end are indicated by the dashed lines (the length of each QA list is fixed --- in reality, QA lists are useful only for much longer positions lists, e.g., ones that do not fit into a single cache line). Let us note that this layout is based directly on the layout that we have devised for another quick-access text-based data structure called a split index~\cite{CislakG15}. Adapted from Wikimedia Commons (author: Jorge Stolfi; available at \url{http://en.wikipedia.org/wiki/File:Hash_table_3_1_1_0_1_0_0_SP.svg}; CC A-SA 3.0).}
    \label{Fig:layout}
\end{figure}

As regards the implementation, the index uses a hash table in order to store (for each selected $q$-gram) the value of the count table and the occurrence positions which are used by the 
$Occ(\cdot)$ function.
Rather than having bitvectors of length $n$ (e.g., as the wavelet tree components), 
as is usually the case for the FM-index,
we explicitly store only the occupied positions (as 32-bit integers), 
since otherwise the space requirements would be prohibitive ($O(n^2 \log n)$ bits).
The search itself is either a linear search if the list size is, roughly speaking, not much greater than one cache line, or otherwise a binary search (BS).
Moreover, in order to speed up this search, we store an additional quick access (QA) list.
This list points to values corresponding to consecutive percentiles on the list --- 
we first perform a linear search on a short QA list, and then jump to 
the specific position on the positions list.
Optimal values such as the length of the QA list and the binary search threshold 
(i.e.~the minimum list length which is required to use the BS instead of a linear search) 
were determined empirically.
The buckets, which store the $q$-grams and pointers to their respective lists, 
have a contiguous layout, as illustrated in Figure~\ref{Fig:layout}.

The experimental results were obtained on a machine equipped with the 
Intel i5-3230M processor running at 2.6\,GHz and 8\,GB of DDR3 memory, 
and the C++ source code was compiled (as a 32-bit version) with 
clang v.~3.6.2-1 and run on the Ubuntu 15.04 OS.
In the following paragraphs, we present the results for the superlinear variant.

As regards the hash function, xxhash (\url{https://code.google.com/p/xxhash/}) was used.
We have considered two pattern splitting schemes, namely one based on the powers of 2,
and the other based on the sequence of Fibonacci numbers.
The results for both cases differ significantly due to different relations between the sizes of stored $q$-grams and specific patterns, still, we can see the characteristic spikes, with the lowest search times reported for pattern lengths that matched the length of one of the stored $q$-grams (i.e.~when it was equal to the power of 2 or to one of the Fibonacci numbers).
The times are given per character and they represent the average times calculated for $10^6$ queries which were extracted from the input text.

\begin{figure}[h!]
    \centering
    \includegraphics[scale=0.55]{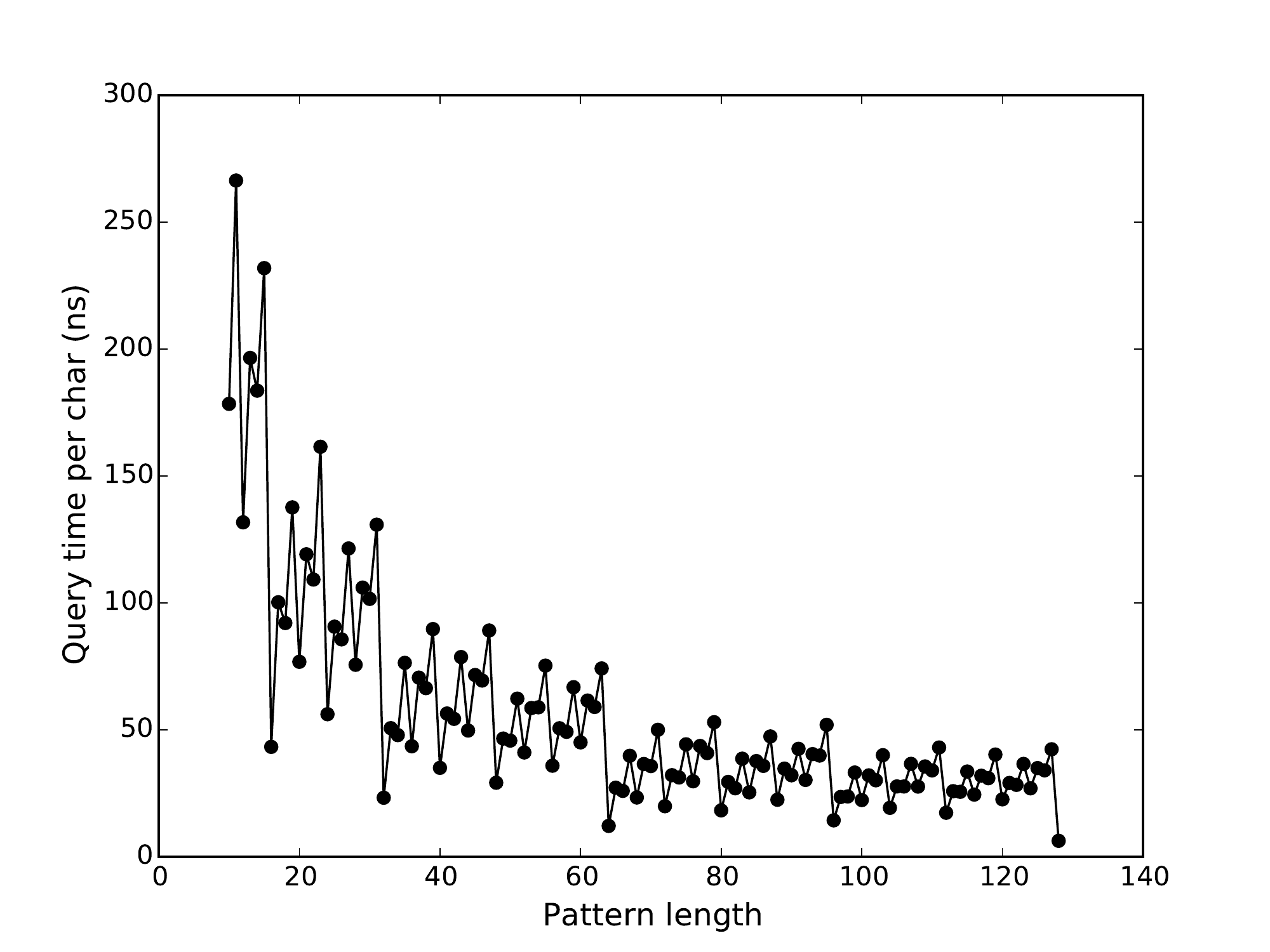}
    \includegraphics[scale=0.55]{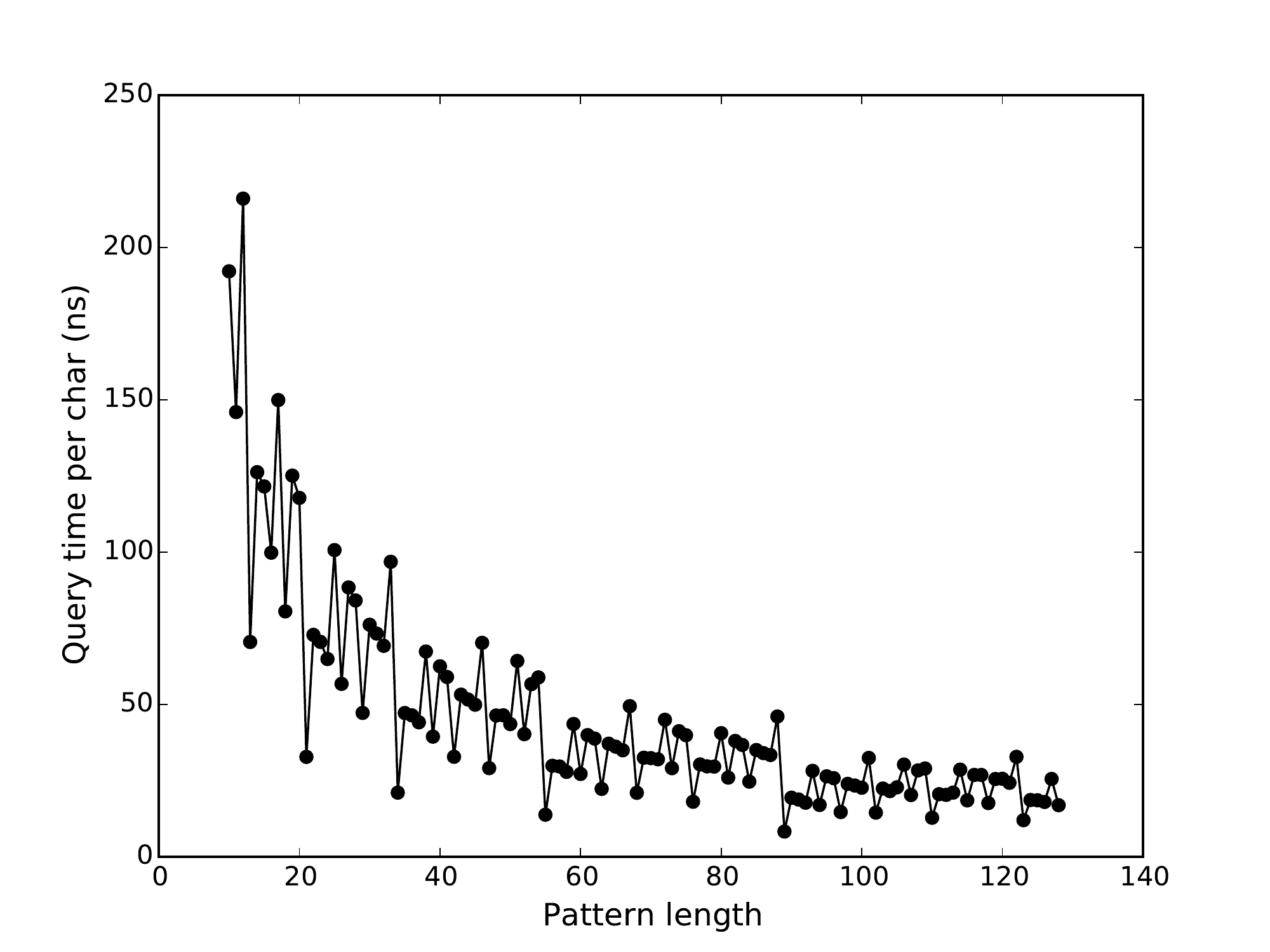}

    \caption{Query time per character vs pattern size (from 10 to 128, inclusive) for the superlinear FM-bloated index for powers of 2 (upper diagram) and Fibonacci numbers (lower diagram) for the English text of size 25\,MB. A hash table with quick access lists was used, which allowed for a slight speedup with respect to a hash table without said lists at the cost of a negligible increase in index size.}
    \label{Fig:patsize}
\end{figure}

In order to evaluate both schemes, we have calculated the total sum of all search times for the pattern lengths from 10 to 128 (both inclusive).
The former turned out to produce a relatively smaller index (around $80n$) with the total time being equal to 283.18\,ms, and the latter produced a relatively bigger index (around $95n$) with the lower total time of 225.91\,ms; consult Figure~\ref{Fig:patsize} for details.

\begin{figure}[h!]
    \centering
    \includegraphics[scale=0.55]{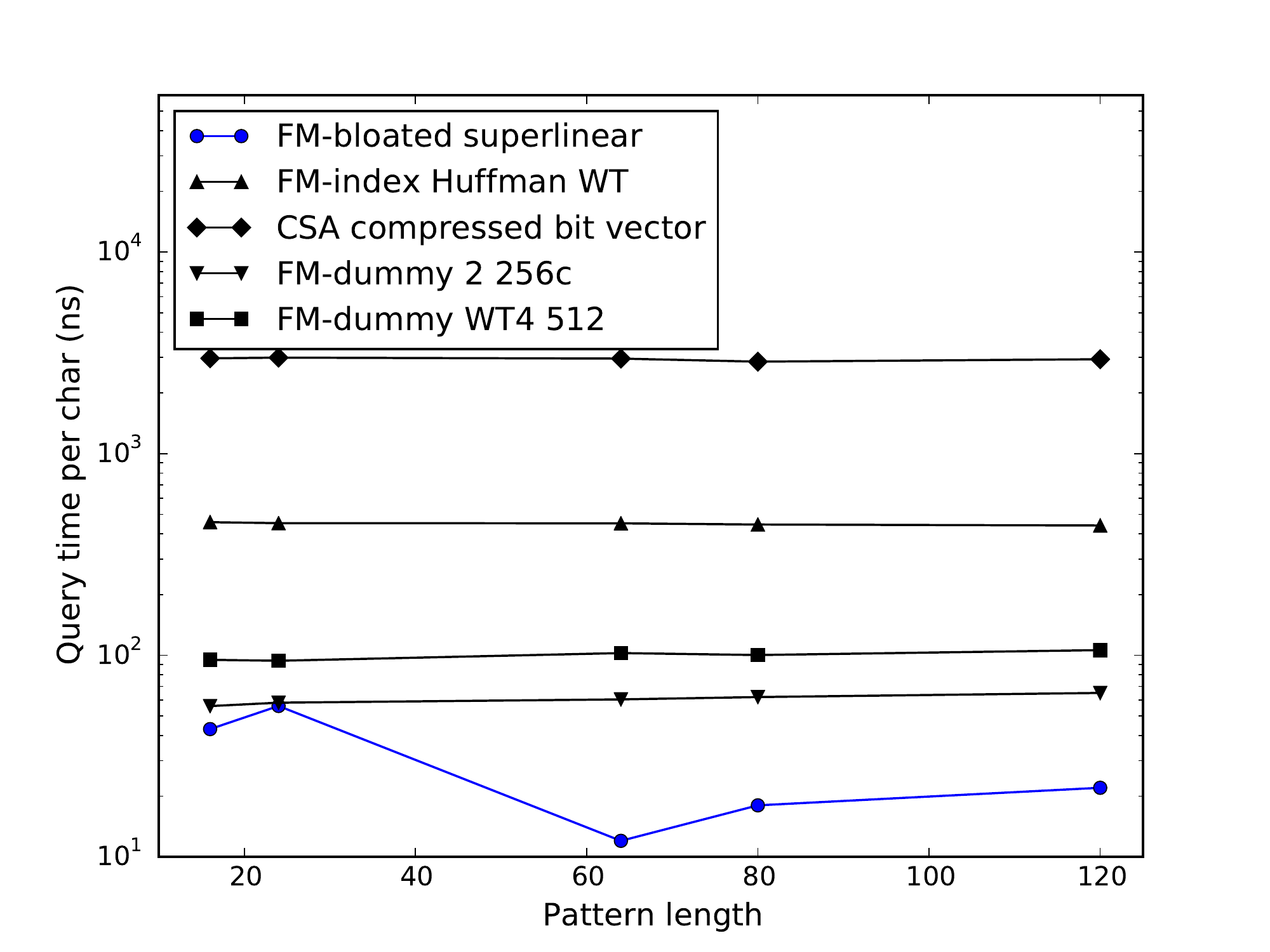}

    \caption{Query time per character vs pattern size (16, 24, 64, 80, and 120) for different methods for the English text of size 25\,MB. The FM-bloated superlinear version with powers of 2 was used. Note the logarithmic y-scale.}
    \label{Fig:comp}
\end{figure}

In Figure~\ref{Fig:comp}, we can see a comparison with other FM-index-based structures.
We used the implementations from the sdsl library (\url{https://github.com/simongog/sdsl-lite}) and the implementations of FM-dummy structures by Grabowski et al.~\cite{GrabowskiRD15} (\url{https://github.com/mranisz/fmdummy/releases/tag/v1.0.0}). 
As regards the space usage, the FM-bloated structure (just as the name suggests) is roughly two orders of magnitude bigger than other indexes (the index size for other methods ranged from approximately $0.6 n$ to $4.25 n$).

\section*{Acknowledgement}
The work was supported by the Polish National Science Centre upon decision 
DEC-2013/09/B/ST6/03117 (the first author).

\bibliographystyle{abbrv}
\bibliography{bloated}

\end{document}